# IMPROVING CNN-BASED STOCK TRADING BY CONSIDERING DATA HETEROGENEITY AND BURST

Keer Yang[1], Guanqun Zhang[2], Chuan Bi[3], Qiang Guan[4], Hailu Xu[5], Shuai Xu[1]

[1] Case Western Reserve University
[2] Nankai University
[3] National Institute of Health
[4] Kent State University
[5] California State University Long Beach

## ABSTRACT

*In recent years, there have been quite a few attempts to apply intelligent techniques to financial trading, i.e., constructing automatic and intelligent trading framework based on historical stock price. Due to the unpredictable, uncertainty and volatile nature of financial market, researchers have also resorted to deep learning to construct the intelligent trading framework. In this paper, we propose to use CNN as the core functionality of such framework, because it is able to learn the spatial dependency (i.e., between rows and columns) of the input data. However, different with existing deep learning-based trading frameworks, we develop novel normalization process to prepare the stock data. In particular, we first empirically observe that the stock data is intrinsically heterogeneous and bursty, and then validate the heterogeneity and burst nature of stock data from a statistical perspective. Next, we design the data normalization method in a way such that the data heterogeneity is preserved and bursty events are suppressed. We verify out developed CNN-based trading framework plus our new normalization method on 29 stocks. Experiment results show that our approach can outperform other comparing approaches.*

## 1. INTRODUCTION

Stock market prediction has gotten a lot of interest in academic and industrial research over the last few decades, thanks to the possibility for big rewards if the stock market moves in the predicted direction. Because investing in the stock market is believed to be more volatile and sophisticated than investing in other markets such as bonds and real estate, the profitability of active stock investing is greatly dependent on one's ability to predict market movement. It is worth noting that two alternative viewpoints on the stock market's predictability have existed for decades, ever since the Efficient-Market Hypothesis (EMH) [1] was first proposed, which states that stocks always trade at their fair value prices, making it impossible for investors to consistently outperform the overall market (often reflected as market indexes). In the latter case, stock market forecasting is regarded as an important aspect of financial time series forecasting because it provides insights into the overall state of the economy as well as market-timing information. For decades, people have attempted to construct autonomous intelligent decision-making models for stock trading by resorting to techniques developed in Statistics [2], machine learning (ML) [3], and artificial intelligence (AI) [4]. Among the vast amount of studies that have been proposed to predict the stock market movement, deep learning (DL) implementations of the financial time series forecasting started to emerge in the past decade, as it is considered as one of the best performing predictors by far, compared to other statistical and classical ML models. At the same time, increasing financial instruments have been created for individual investors and traders, which also make the problem of algorithmic/intelligent autonomous trading possible and appealing [5].

Deep learning-based methods have shown state-of-the-art performance on various prediction and classification problems, and they beat traditional techniques such as regressions and support vector machines [6]. As a result, researchers have attempted to develop deep learning approaches to solve their domain-specific problems, such as monitoring the health of a bridge [7], smart agriculture [8], adding sound to silent movies [9], and economical situation prediction [10]. Essentially, mostly of the aforementioned applications are successful, because deep learning-based methods are able to discover stealthy mapping between the input data and the desired output labels. To be more specific, these applications adopt deep learning to analyze the time series data generated under various scenarios by extracting hidden features and constructing predictive models.

In this work, we investigate the problem of constructing an autonomous trading framework by using deep learning. Although a few works have addressed the similar problem by adopting recurrent neural networks (RNN) [11], [12], convoluted neural networks (CNN) [13]–[15], and long short term memory (LSTM) [16]–[18], one important aspect stock data, the nature of data heterogeneity and burstiness (formally discussed in Section IV) was ignored, and usually result in sub-optimal trading strategy. In this paper, we constructed a CNN model that takes data heterogeneity and burstiness into consideration to automatically and adaptively learn spatial hierarchies of features through backpropagations of the stock market data. The reason we choose CNN is because it considers spatial correlation of data via spatial convolutions. In addition, different CNN-based architectures for image classifications [19], [20] and CNN-based stock data prediction [15] were proposed in previous studies. In this research, we propose a novel approach to pre-process (i.e., data normalization) the collected data to preserve the heterogeneity of stock data and at the same time to suppress the bursty events observed in the data. The data collection procedure is deferred to Section III.

After data collection, we observe that the collected stock data is heterogeneous in feature dimension but homogeneous in time dimension by comparing with image inputs (the most common inputs for CNN models). Then, we validate the observed heterogeneity using various Statistical metrics, e.g., entropy and Chi-square test, which further inspire us to preserve the data heterogeneity by conducting data normalization row by row (instead of sample by sample adopted by conventional CNN models). Furthermore, we notice that the stock data also has the same properties as other social dynamics, i.e., the distribution for the magnitude of social dynamics records is right-skewed; it is typically small but can be occasionally very large. Thus, we propose an instantiation of the rowwise normalization by re-scaling the stock data using a logarithm function, because it helps suppress the abnormal records (that has large value) and widely spread normal records on log-scale.

We experimentally show that our developed data normalization method better preserves the data heterogeneity, controls the impact of abnormal records, and thus, better suits the need of designing a CNN-based autonomous trading framework. We empirical validate this by comparing with the frameworks constructed using various models, i.e., RNN, regression, and CNN with conventional normalization methods. Experiment results show that considering a long-term investigation (e.g., 10 years), the trading policy achieved by our framework can lead to much higher total assets for most of the considered stocks.

The contributions of this paper are listed as follows.

- We empirically observe that the stock data is intrinsically heterogeneous (given different financial indicators) and bursty. We also statistically validate the heterogeneity and burst nature of stock data .
- We propose new data normalization method specifically for the stock data in order to preserve the heterogeneity and suppress bursty events (i.e., outlier events with large magnitudes and rarely occur).
- We develop a CNN-based automatic trading framework which consumes normalized stock data and outputs trading strategies, i.e., buy, sell, or hold.
- We experimentally compare our developed CNN-based trading framework with LSTM- and regression- based frameworks on 29 common stocks (e.g., AAPL, MSFT, AMD, etc). Experiment results show that our framework not only achieves better computational results (e.g., higher precision or $F_1$ score), but can also lead to higher simulated asset in the long-term trading.

The rest of this paper are organized as follows. In Section II, we review related works, which is followed by preliminaries on CNN and stock data collection in Section III. In Section IV, we develop our proposed data normalization technique to preserve data heterogeneity. In Section V, we conduct extensive experiment by consider various stocks and compare with other state-of-the-art models. Finally, Section VI concludes the paper.

## 2. RELATED WORK

Although deep learning have been widely applied in the fields of image recognition, autonomous driving, computer vision, there are only a few attempts of implementing deep neural networks for financial problems. Ding et al. [6] develop event-driven based deep learning method for stock market prediction. In particular, they first extract events from text of financial news and represent the events as dense vectors, based on which a neural network is trained to model both short- and long-term impacts of events on stock price movements. Fischer et al. [16] used LSTM to construct an automatic stock trading platform and compared with other memory-free models, e.g., random forest and logistic regression classifier. Krauss et al. [11] conduct an extensive comparison between trading models obtained using deep neural networks, gradient-boosted-trees, random forests, and ensembles of these models. They find that as long as neural networks are involved, the performance of stock prediction is significantly improved. Yoshihara et al. [21] propose to use RNN to predict the trend of stock prices on the Nikkei Stock Exchange using news events. They also compared their method with support vector machine. Sezer et al. [15] propose to use CNN on stock market prediction by extracting features using financial indicators and shape the features as 2D images. Since we also construct the stock prediction model using CNN, we will also compare with [15] in the experiments. Finally, please refer to Langkvist et al. [22] for a survey about using various deep learning models for stock prediction.

Unfortunately, none of the above-mentioned works have explicitly consider the heterogeneity and burst nature of stock data; they directly feed in the extracted financial data into various learning models. In this paper, we to preserve the data heterogeneity and limit the impact of abnormal (bursty) data records, we proposed a novel data normalization method.

## 3. PRELIMINARIES

In this section, we first discuss the process of stock data collection and then review the structure of CNN.

### 3.1. Data Collection

The core idea of collecting stock data as 2D images can be summarized in the following two steps.

- Technical indicators calculations. For each trading day, we choose 15 different period lengths, calculate 15 technical indicators, and then, stack the results into a 15×15 matrix (i.e., image), where each row corresponds to one specific technical indicator.
- Image labelling. We label each of the obtained image as buy, sell, or hold (i.e., 0, 1, or 2) based on certain metrics (e.g., the closing price).

In particular, we take the simple moving average (SMA) indicator as an example and consider the trading day on March 20th, 2019. We set the window length as $l \in \{6,7,...,20\}$. Then, the SMA value of a specific stock considering consecutive $l$ trading days starting from March 20th, 2019 is $SMA(d,l) = \frac{1}{l}\sum_{i=0}^{l-1} p(i+1)$, where $d$ indicates March 20th, 2019, and $p(i)$ represents the close price of that stock on day $i$. After calculating the SMA indicator for the considered window length, we collect the results as a row vector. Then, by repeating the same process for other indicators and stacking all resultant row vectors as a 2D matrix, we can achieve a stock image on $d$ = March 20th, 2019, i.e., $x(d) \in R^{15 \times 15}$. In this paper, we achieve a fair comparison, we adopt the same technical indicators as used in [15].

Then, we label $x(d)$ as buy, sell, or hold by considering its close price, $p(d)$, in the longest window length (i.e., $l$ = 20). In particular, if $p(d)$ is lower than the first quartile (i.e., 25th percentile) of the 20 consecutive trading days, we label $x(d)$ as buy, and if $p(d)$ is higher than the third quartile (i.e., 75th percentile), we label $x(d)$ as sell. Otherwise, we label $x(d)$ as hold. It is noteworthy that our labelling method is different with the existing ones, e.g., [15], which just labels $x(d)$ as buy (or sell) if $p(d)$ is the lowest (or highest) in the considered window length.

### 3.2. Architecture of CNN

In particular, our CNN model is composed of the following layers.

- The input layer which consumes the constructed stock data images of size $15 \times 15$.
- Two convolutional layers, where the first convolutional layer is composed of 32 filters with size 4×4 and the second one consists of 64 filters with size $4 \times 4$. Convolutional layers are used to extract the various features from the input stock images.
- A max pooling layer decreases the size of the output of the previous convolutional layer in order to reduce the computational cost. In this layer, we only keep the largest element in each $2 \times 2$ block.
- The fully connected layer maps the output of the max-pooling layer into a $128 \times 1$ vector. In this layer, the input image from the previous layers are flattened, the flattened results is $1024 \times 1$), and then mapped to the desired dimension.
- Dropout and output layer. Finally, the $128 \times 1$ vector is mapped to the outputs (e.g., buy, sell, and hold) via fully connection. We also perform dropout in this process by randomly setting some of the connection weights to 0 in order to control overfitting. In this paper, the dropout ratio is 0.3.

Another important component of CNN is the activation functions, which are adopted to learn and approximate the continuous and complex relationship between variables of neural network. Technically, they decide which information of the CNN model should fire in the forward direction. They also add on non-linearity to the network by involving non-linear transformations. There are several commonly used activation functions such as the ReLU, softmax, tanh and the sigmoid. Each of these functions have a specific usage. In this paper, we consider the sigmoid activation function, which is bounded between 0 and 1.

## 4. OUR METHODS

In this section, we first discuss the process of stock data collection and then review the structure of CNN.

### 4.1. Observation of Heterogeneity Across Features

Different with 2D images, which are usually homogeneous along the two dimensions (i..e, a specific pixel is similar to its surrounding pixels), the constructed 2D stock data has inherent heterogeneity in the feature dimension (due to the choice of different economical indicators) but has homogeneity in the time dimension. The heterogeneity can be interpreted as each feature vector obtained using a certain economical indicator demonstrates a specific financial pattern and no one is an exact copy of another. To validate this, we compare the 2D stock data with real 2D facial images in Figure 1. Clearly, as visualized in Figure 1 (b), each row (i.e., each feature vector) of the stock data are unique; there is less similarity across rows. In contrast, the facial images in Figure 1 (a) presents homogeneity across both dimension, e.g., the pixel intensity are similar in the area of a human cheek. Besides, we also show the value of entropy on top of each visualized data in Figure 1. In particular, entropy is a statistical measure of randomness that can be used to characterize the texture of the input image, and the higher the entropy the more heterogeneous. We observe that the entropy of 2D stock data are much higher (larger than ×15 the entropy of the grey-scale facial images), which suggests that the constructed 2D stock data has inherent heterogeneity in the feature dimension.

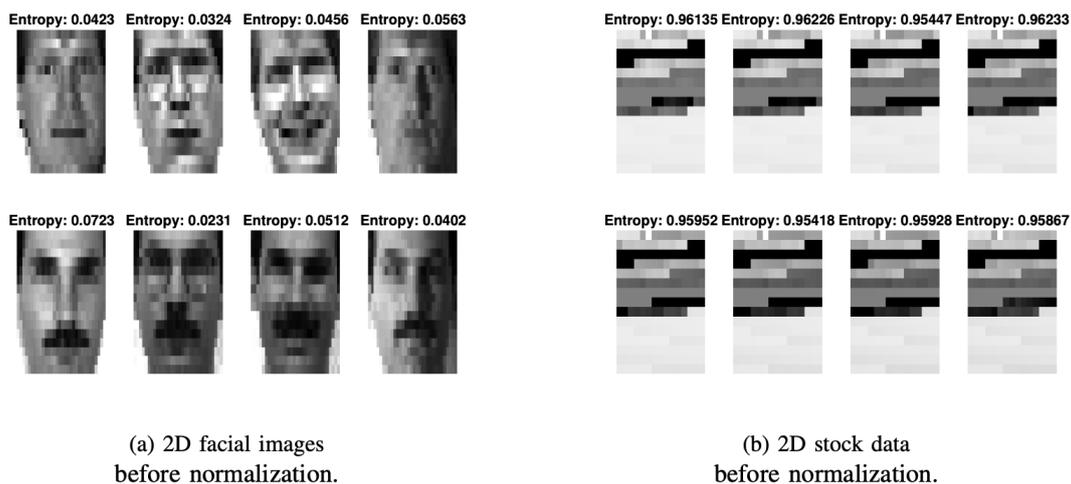

(a) 2D facial images
before normalization.

(b) 2D stock data
before normalization.

Fig. 1: Visualization of the original grey-scale 2D face images and 2D stock data, where the entropy is computed before they are normalized.

## 4.2. Validation of Data Heterogeneity via Chi-square Test

In this section, we validate the observation of data heterogeneity by means of Chi-square test. In particular, the null-hypothesis $H_0$ and alternative-hypothesis $H_a$ are defined as

$H_0$: observation across different indicators are similar,
$H_a$: observation across different indicators not similar,

and the choose the significant level as 1%. Still, we take the facial images and stock data images in Figure 1 as an example. To conduct the Chi-square test, we let the first row in each image (individual's face or a stock data) as the expected values, and the other rest rows in that image as observations. Since there are 15 observations in each row, we have the degree of freedom as 14, and the significant value is 29.141. We have the Chi-square statistics of between the observation in the $i$th row ($i \in \{2,3,...,15\}$) and the first row is calculated as

$$\chi^2(i) = \Sigma(x_{ij} - x_{1j})^2 / x_{1j}, \; i \in \{2,3,...,15\}.$$

As long as $\chi^2(i)$ is larger than 29.141, the null-hypothesis will be rejected, i.e., the observations across different indicators are not similar. In Figure 2 we provide the bar plot of the chi-square statistics for the 2nd to 15th rows. The black horizontal line represents the value of 29.141. Clearly, all chi-square statistics values are higher than 29.141, which suggests that the mull-hypothesis is rejected and different financial indicators are heterogeneous.

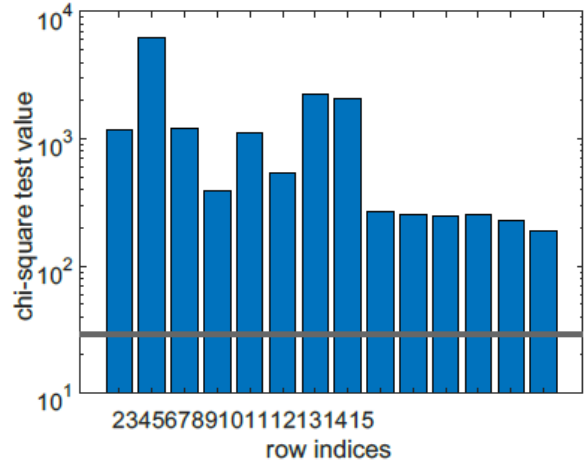

Fig. 2: Chi-square statistics of 2nd to 15th rows.

## 4.3. Row-by-Row Normalization Preserve Heterogeneity

The common data normalization methods in the literature are not suitable for CNN-based stock data analysis. This is because they are applied on the entire input data instead of row by row (or entry by entry). For example, one common practice is the min-max normalization, i.e.,

$$\tilde{x} = \frac{x - \min(x)}{\max(x) - \min(x)}. \tag{1}$$

where $x$ represents a whole image (or batch of images).

In order to preserve heterogeneity of the stock data, we propose to conduct data normalization row by row (since each row corresponds to one specific financial indicator). Thus, inspired by (1), we have

$$\tilde{x}_i = \frac{g(x_i) - \min(g(x_i))}{\max(g(x_i)) - \min(g(x_i))}, \quad (2)$$

where $x_i$ represents one row in the constructed 2D stock data and function $g(\cdot)$ is used to suppress abnormal records but distinguish normal records. We will elaborate $g(\cdot)$ in the next section.

### 4.4. Normality versus Burstiness

Quite a few studies on social dynamics have identified the burstiness in social events [23]–[25], i.e., the distribution for the magnitude of the social dynamics is right-skewed; it is typically small but can be occasionally very large. Similar phenomena has also been reported when analyze signal in the Fourier domain [26]. We observe that the dynamics of finical indicators also show the similar pattern, i.e., abnormality in stock data (finical indicators with extreme large values) is rare and has a burst nature, i.e., it happens very low probability (its frequency is low on a histogram plot). Whereas, normal data records (finical indicators with small to medium value) are very common. We take the CSCO stock as an example, plot the data records along with the histograms of indicators Relative Strength Index (RSI), Williams%R, and Exponential Moving Average (EMA) for 447 days in Figure 3. In particular, Figure 3 (a), (b) and (c) show the plot of the time series, and clearly, the abnormal indicator records all demonstrate a bursty nature, i.e., short duration but higher value. Figure 3 (d), (e) and (f) show the histogram plots, where the lowest bars denotes the number of occurrence of rare (bursty) events. Actually, Figure 1 (b) also shows the similar observation, i.e., for each chosen indicator, the value is more constant for consecutive slide windows, whereas the abnormal value rarely appears.

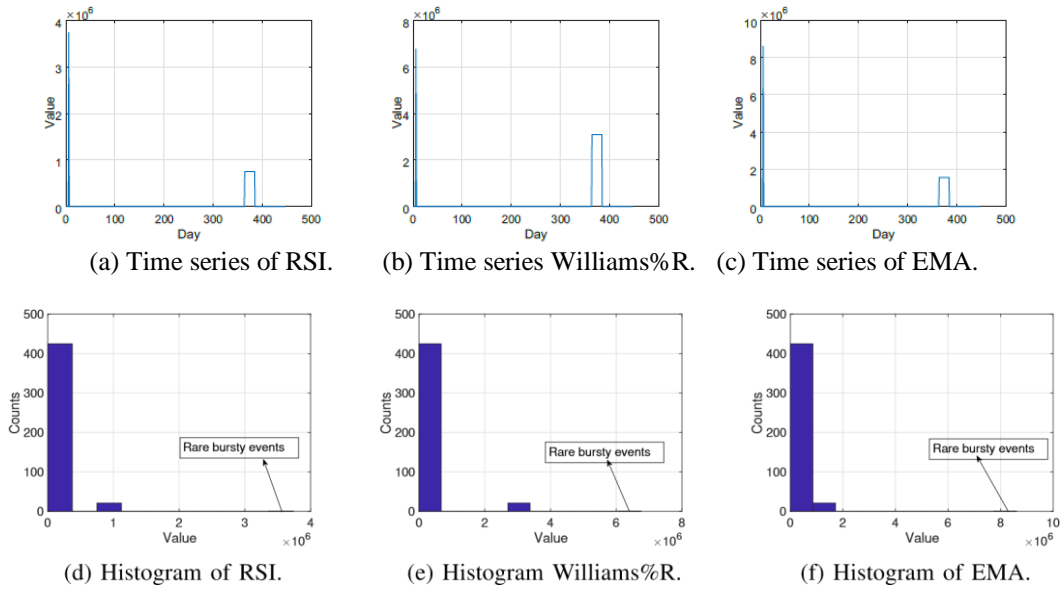

(a) Time series of RSI.   (b) Time series Williams%R.   (c) Time series of EMA.

(d) Histogram of RSI.   (e) Histogram Williams%R.   (f) Histogram of EMA.

Fig. 3: Time series and histogram of Relative Strength Index (RSI), Williams %R, and Exponential Moving Average (EMA), of CSCO stock for 447 days.

It is been widely acknowledged that rare events will compromise the prediction of social trends and patterns [27], [28], because they only have short duration and cannot describe the long-term characteristics of social dynamics. As a matter of fact, bursty events in financial indicators are usually caused by events like war, nature disaster, and economic crisis [29], which are also rare events and cannot be used to guide long-term stock trading. As a result, we want to suppress the value of rare events and at the same time separate (i.e., distinguish) the value of normal indicator records. To do so, we use a monotonically increasing concave function $g(\cdot)$ to re-scale the original data, and let $g(\cdot)$ be the logarithmic function. Then, to ensure non-negative, we essentially instantiate (2) as

$$\tilde{x}_i = \frac{\log(|x_i|) - \min\big(\log(|x_i|)\big)}{\max\big(\log(|x_i|)\big) - \min\big(\log(|x_i|)\big)}, \qquad (3)$$

where $|\cdot|$ takes care of the negative data records.

In Figure 4 (a) and (b), respectively, we visualize the original stock data after min-max normalization (i.e., (1)) and after the proposed normalization (i.e., 3). On top of each normalized data, we also show the new entropy value.

Clearly, for each 2D stock data, our proposed normalization leads to a higher entropy value, which suggests that we can promote the data heterogeneity compared with min-max normalization.

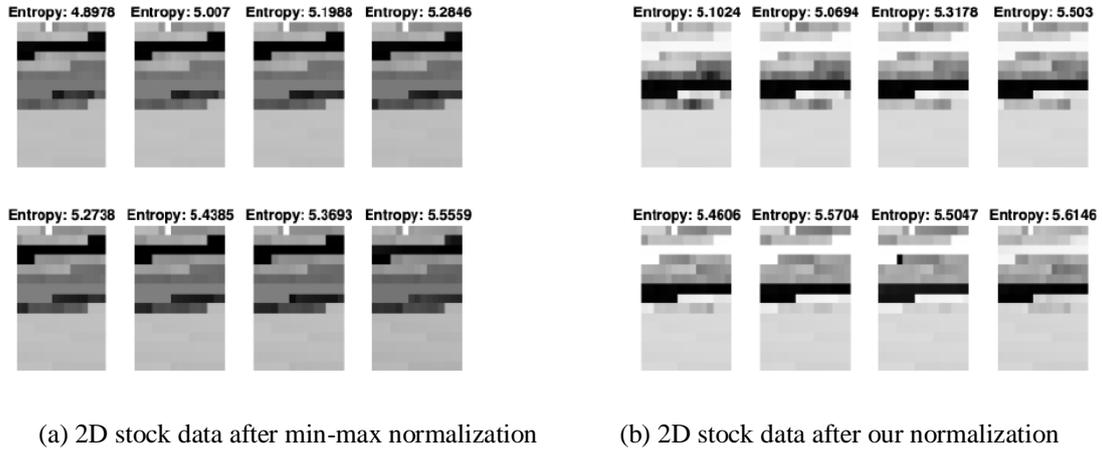

(a) 2D stock data after min-max normalization  (b) 2D stock data after our normalization

Fig. 4: Visualization of the 2D stock data after min-max normalization (i.e., (1)) and after the proposed normalization (i.e., equation (3)).

## 5. EXPERIMENTS

In this section, we conduct extensive experiments to validate our proposed autonomous trading framework and compare with other state-of-the-art approaches, i.e., the trading framework based on CNN with min-max normalization [15], LSTM [30], and linear regression [31].

In particular, LSTM is a variety of recurrent neural networks (RNNs) that are able to learn long-term dependencies of the input data. Compared with a normal artificial neural network, RNN tries to exploit historical information. To be more specific, in an RNN, the output will not only depend on the current input, but also depend on inputs in the history. LSTM networks usually have the same network structure with RNNs, but they use a different way (called memory cell). In this paper, we adopt the model proposed in [30] to construct a LSTM-based trading framework. Besides, we also consider a simple model built on linear regression [31] with Lasso regression [32], i.e., the stock data is mapped to the output labels under linear transformation, and the transformation operator is subject to sparsity constraint.

In the experiments, we adopt two different evaluation criteria to evaluate the automatic trading performance, i) the performance of computational model which evaluates the prediction accuracy of various learning models, i.e., how well different classifiers distinguish between buy, hold and sell labels for a specific stock on a given day, and ii) financial performance that evaluates various models by implementing the real world financial scenario, i.e., the stocks are bought, sold or held on each day according to the predicted labels.

In this experiments, we consider 29 stocks (they are listed in the first column of Table II) in the period of 2005-2021. After pre-processing the stock data into 2D images (discussed in Section III-A), we use the data from 2005 to 2015 as the training dataset and those from 2016-2021 as the testing dataset.

## 5.1. Comparison of Computational Performance under Different Normalization Methods

Here, we first report the precision, recall, and $F1$ score of representative stocks obtained using our proposed normalization technique (to preserve data heterogeneity and suppress burst events) followed by CNN. To be more specific, given a particular stock prediction, the precision of the "Buy" class is the ratio between the number of stock images correctly predicted as "Buy" and the number of all predicted 'Buy''. Whereas, the recall of the "Buy" label is the ratio between the number of stock images correctly predicted as "Buy" and the number of stock images that are actually labeled as "Buy". Then $F1$ score for "Buy" is defined as $2\frac{\text{precision}\times\text{recall}}{\text{precision}+\text{recall}}$. In Table 2(a), we list the results for stocks WMT, MMM, AAPL, and AXP. In Table 2(b), we show such results obtained by [15], which is also CNN based framework where data normalization is conducted using min-max.

From Table 2, we observe that the precision values for "Buy" and "Sell" are usually lowered than that of "Hold", whereas the precision values for "Buy" and "Sell" are usually higher than that of "Hold". This is mainly because of the fact that "Buy" and "Sell" labeled stock data are much less frequent than "Hold" points, thus, it is difficult for CNN models to learn the patterns of these stock data without compromising fitting the distribution of the input data. Similar phenomenon has also been reported in [15]. However, we are able to achieve higher $F1$ score than [15], because we have made the training and testing data more balanced by considering the percentile of the stock price for each day in consecutive trading days.

TABLE I: Precision, recall, and $F1$ score for WMT, MMM, AAPL, and AXP achieved by (a) our proposed normalization method followed by CNN, and (b) min-max normalization method followed by CNN, i.e., [15]

|  |  | Buy | Sell | Hold |
|---|---|---|---|---|
| WMT | precision | 0.34 | 0.21 | 0.82 |
|  | recall | 0.84 | 0.85 | 0.56 |
|  | $F1$ score | 0.48 | 0.37 | 0.66 |
| MMM | precision | 0.57 | 0.33 | 0.72 |
|  | recall | 0.71 | 0.73 | 0.43 |
|  | $F1$ score | 0.63 | 0.45 | 0.54 |
| AAPL | precision | 0.41 | 0.38 | 0.68 |
|  | recall | 0.81 | 0.74 | 0.45 |
|  | $F1$ score | 0.54 | 0.50 | 0.54 |
| AXP | precision | 0.37 | 0.41 | 0.77 |
|  | recall | 0.77 | 0.63 | 0.51 |
|  | $F1$ score | 0.50 | 0.56 | 0.61 |

(a) Our normalization method followed by CNN.

|  |  | Buy | Sell | Hold |
|---|---|---|---|---|
| WMT | precision | 0.27 | 0.19 | 0.72 |
|  | recall | 0.74 | 0.61 | 0.46 |
|  | $F1$ score | 0.40 | 0.29 | 0.56 |
| MMM | precision | 0.34 | 0.25 | 0.68 |
|  | recall | 0.68 | 0.71 | 0.39 |
|  | $F1$ score | 0.45 | 0.37 | 0.50 |
| AAPL | precision | 0.38 | 0.36 | 0.66 |
|  | recall | 0.76 | 0.69 | 0.55 |
|  | $F1$ score | 0.51 | 0.47 | 0.60 |
| AXP | precision | 0.33 | 0.40 | 0.76 |
|  | recall | 0.68 | 0.65 | 0.57 |
|  | $F1$ score | 0.44 | 0.50 | 0.53 |

(b) Min-Max normalization followed by CNN.

## 5.2. Financial Performance

In this section, we compare our proposed autonomous trading framework with other approaches by simulating transactions. In particular, for each stock data in the testing dataset, we perform buy, sell, or hold according to the predicted labels obtained by various models. In particular, we start with 10,000$ cash and the amount of stocks that worth 10,000$ (which gives 20,000$ initial findings). For each day from 2016-2021, if the predicted label is "Buy", we will purchase that stock with 50% of the cash, if the predicted label is "Sell", we will sell that 50% shares of that stock, and if the predicted label is "Hold", we will not make any transaction.

In Table II, we record the multi-class prediction accuracy and annualized returns of different models for all considered 29 stocks. We define the return of a trading strategy as the increase of the total capital (cash and stock) at the end of 2021. In Table II, the bold entries correspond to the highest return obtained by all frameworks. Clearly, for most of the considered stocks, we can achieve the highest return in 5 years. It validates that by taking account of the data heterogeneity and suppressing the burst events (data records with extreme large magnitude), we can achieve more robust automatic stock trading framework.

Furthermore, we also plot the total assets versus time in Figure 5, where the red, blue, green, and black curves, respectively, represent the assets achieved by our framework, , LSTM , and the regression method . Clearly, we can see for most of the stocks, our framework (i.e., red curves) can lead to higher assets. We also observe that although more complex, the LSTM-based framework does not have better results, and for some stocks (e.g., CVX and CSCO) it cannot make any profit. This is because, the constructed stock image data is small and LSTM networks can easily overfit the training dataset. Another drawback of LSTM-based framework is that it consumes much more time to train. On the other hand, the regression-based method also has poor performance due to its model simplicity which ignores the non-linear relationship between the input stock data images and the output labels.

| Stock metrics | CNN+log-Norm | | CNN+Min-max Nrom | | LSTM | | Regression | |
|---|---|---|---|---|---|---|---|---|
| | Accuracy | Return | Accuracy | Return | Accuracy | Return | Accuracy | Return |
| MMM | 0.51 | +1.5% | 0.57 | **13%** | 0.43 | -1.3% | 0.37 | -12% |
| AXP | 0.72 | **+81%** | 0.66 | +23% | 0.52 | +26% | 0.55 | +33% |
| AAPL | 0.53 | +50% | 0.32 | 0% | 0.34 | 0% | 0.76 | **+254%** |
| WMT | 0.71 | **+113%** | 0.59 | +65% | 0.57 | +67% | 0.52 | +61% |
| BA | 0.68 | +66% | 0.48 | -22% | 0.54 | **+67%** | 0.39 | -27% |
| CAT | 0.68 | **+69%** | 0.43 | -19% | 0.52 | +68% | 0.33 | -22% |
| CVX | 0.83 | **+35%** | 0.74 | +24% | 0.51 | +0% | 0.49 | -32% |
| CSCO | 0.67 | **+39%** | 0.61 | +20% | 0.44 | +3.5% | 0.42 | +11% |
| KO | 0.65 | +19% | 0.47 | -5% | 0.56 | **+21%** | 0.59 | 0% |
| DIS | 0.78 | **+76%** | 0.66 | +68% | 0.62 | +43% | 0.55 | +62% |
| XOM | 0.78 | **+24%** | 0.65 | +22% | 0.52 | 0% | 0.50 | -47% |
| GE | 0.71 | **-26%** | 0.65 | -33% | 0.55 | -61% | 0.41 | -35% |
| GS | 0.72 | **+40%** | 0.45 | +13% | 0.52 | 0% | 0.45 | +37% |
| HD | 0.83 | **+115%** | 0.70 | +75% | 0.60 | +68% | 0.62 | +74% |
| IBM | 0.68 | **+32%** | 0.66 | +29% | 0.50 | 0% | 0.47 | +19% |
| INTC | 0.77 | **+82%** | 0.71 | +74% | 0.50 | +69% | 0.39 | +79% |
| JNJ | 0.78 | **+91%** | 0.64 | +41% | 0.52 | +88% | 0.59 | -11% |
| JPM | 0.73 | **+61%** | 0.69 | +38% | 0.53 | +11% | 0.55 | +59% |
| MCD | 0.82 | **+85%** | 0.61 | +83% | 0.65 | +52% | 0.53 | +1% |
| MSFT | 0.81 | **+322%** | 0.67 | +107% | 0.34 | +39% | 0.43 | +115% |
| NKE | 0.83 | **+138%** | 0.65 | +28% | 0.60 | +48% | 0.54 | +13% |
| PFE | 0.71 | **+31%** | 0.65 | +3% | 0.45 | +13% | 0.50 | +25% |
| PG | 0.81 | **+65%** | 0.63 | +41% | 0.62 | 0% | 0.52 | 0% |
| TRV | 0.70 | **+32%** | 0.70 | +18% | 0.49 | +17% | 0.39 | -11% |
| UNH | 0.79 | **+165%** | 0.62 | +113% | 0.30 | +12% | 0.38 | +100% |
| VZ | 0.76 | **+34%** | 0.62 | +23% | 0.52 | +15% | 0.38 | 0% |

TABLE II: Comparison of prediction accuracy and annualized returns of different models.

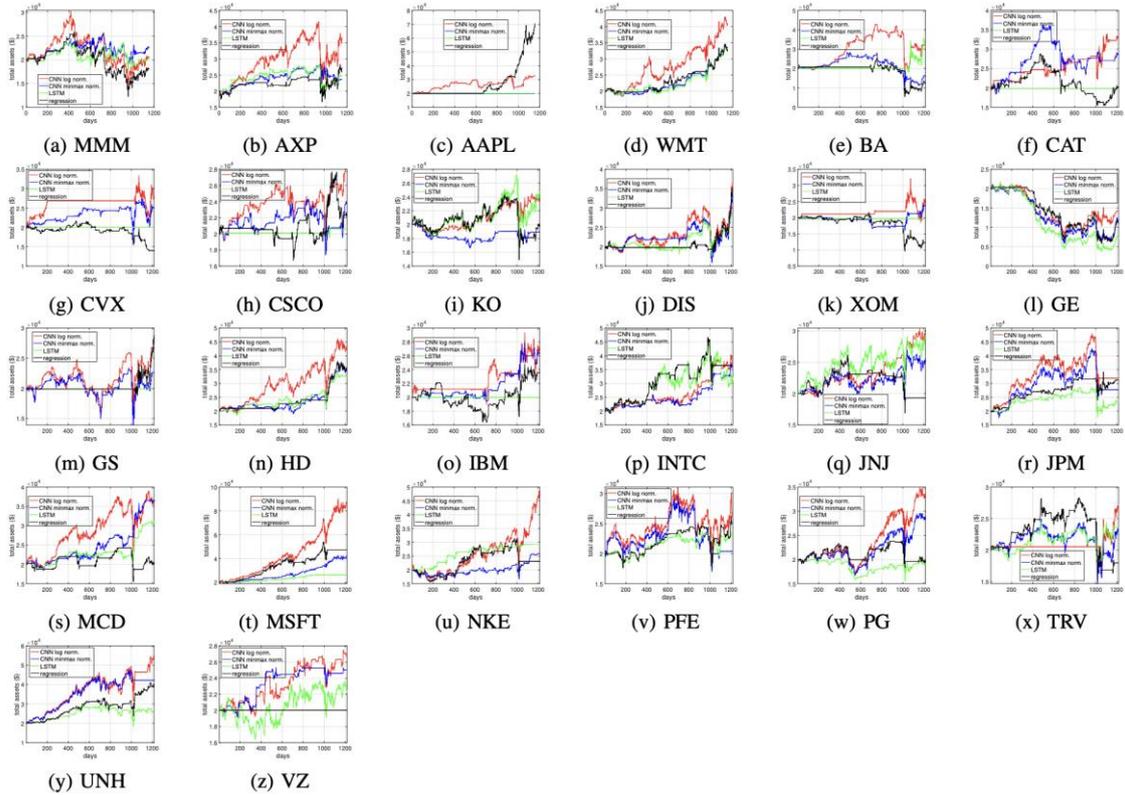

Fig. 5: Total assets versus time of all stocks (part 1) in 5 years achieved by our framework, CNN with min-max normalization [15], LSTM [30], and regression [31].

## 6. CONCLUSIONS

In this paper, we utilized CNN to construct a automatic stock trading framework. We first analyzed financial time series calculated using different financial indicators and converted them into 2-D images. Then, we observe that there are intrinsic data heterogeneity and burst in the obtained stock data, which motivates us to develop a novel data normalization method in order to preserve such heterogeneity and control the impact of abnormal data entries. We verified out developed CNN-based trading framework along with the developed data normalization method using 29 stocks and compare with the LSTM- and regression- based frameworks. Experiments show that our framework.